# Control Industrial Automation System with Large Language Model Agents

Yuchen Xia, Nasser Jazdi, Jize Zhang, Chaitanya Shah, Michael Weyrich
Institute of Industrial Automation and Software Engineering
University of Stuttgart
Stuttgart, Germany
yuchen.xia@ias.uni-stuttgart.de, nasser.jazdi@ias.uni-stuttgart.de, st171260@stud.uni-stuttgart.de, st181599@stud.uni-stuttgart.de,
michael.weyrich@ias.uni-stuttgart.de

***Abstract*— Traditional industrial automation systems require specialized expertise to operate and complex reprogramming to adapt to new processes. Large language models offer the intelligence to make them more flexible and easier to use. However, LLMs' application in industrial settings is underexplored. This paper introduces a framework for integrating LLMs to achieve end-to-end control of industrial automation systems. At the core of the framework are an agent system designed for industrial tasks, a structured prompting method, and an event-driven information modeling mechanism that provides real-time data for LLM inference. The framework supplies LLMs with real-time events on different context semantic levels, allowing them to interpret the information, generate production plans, and control operations on the automation system. It also supports structured dataset creation for fine-tuning on this downstream application of LLMs. Our contribution includes a formal system design, proof-of-concept implementation, and a method for generating task-specific datasets for LLM fine-tuning and testing. This approach enables a more adaptive automation system that can respond to spontaneous events, while allowing easier operation and configuration through natural language for more intuitive human-machine interaction. We provide demo videos and detailed data on GitHub: https://github.com/YuchenXia/LLM4IAS.***

***Keywords*— *Large Language Model, Industrial Automation System, Digital Twin, Intelligent Robotics, Multi-Agent System***

## I. Introduction

Traditional industrial automation systems are rigid, requiring specialized expertise for any modification or reconfiguration. For instance, when the system needs to be adapted to produce new product variants or execute different operations, significant effort is required to design and implement the necessary changes. This process is often hampered by several challenges, including the need for an in-depth understanding of the complicated equipment and the time-consuming effort in translating user requirements into executable programs. These factors contribute to delays and increased costs, with reconfiguration often constrained by knowledge barriers, the intricate nature of reprogramming tasks, and possibly inefficient communication between user and programmer. As a result, traditional industrial automation systems are not only inflexible but also costly and time-inefficient when adapting to new demands[1].

Large language models offer transformative potential in industrial automation. They can perform reasoning based on the knowledge internalized during pre-training, interpret both general and domain-specific language, and generate on-demand responses to varied inputs. While LLMs have demonstrated their utility in general chatbot applications[2], their tailored application in industrial contexts remains underexplored. The challenge lies in effectively adapting these capabilities to deliver tangible value in the industrial domain. A structured approach is required to link the digital functionality of LLMs with the physical realm of industrial automation.

In this paper, we introduce a novel framework for controlling and configuring industrial automation equipment using large language models, enabling more flexible, intuitive and knowledge-informed automation systems. As our contribution, this framework includes:

- An integral system design for applying large language models in industrial automation.
- A proof-of-concept implementation on a physical production system with quantitative evaluation.
- A systematic approach for creating datasets for fine-tuning LLMs to adapt a general pre-trained model for this specific industrial application.

As a result, we present an LLM-controlled automation system that can interpret user tasks specified in natural language, generate a production plan, and execute operations on the physical shop floor. The configuration of control logic is enabled by prompt design, and the application-specific adaptation of the LLMs is enabled by supervised fine-tuning on data collected during system operation.

## II. Requirements and System Setup

This section introduces the required technological foundation in typical industrial automation domain that enables the downstream application of LLM. Based on this foundation, we establish a system depicted in Figure *1*.

### *A. Interoperability*

Interoperability is a fundamental prerequisite for implementing intelligent systems. This concept involves two key aspects: synchronized data acquisition and unified control interface.

PLCs are commonly central to industrial automation, serving as nodes for field data collection and control point exposure. Building on this, OPC UA[3] enables seamless interoperability by providing a unified interface to connect PLCs with higher-level systems across various devices and platforms. In robotics and automation systems that utilize ROS, unified data and control interfaces can be established via ROS communication mechanisms. For proprietary automation modules, communication can be achieved through industrial Ethernet TCP/IP.

Utilizing these technology stacks, various automation modules can be digitally integrated through a unified data and control interface. This integration facilitates the creation of a cyber environment, providing access to the physical system, also referred to as a Cyber-Physical System[4].

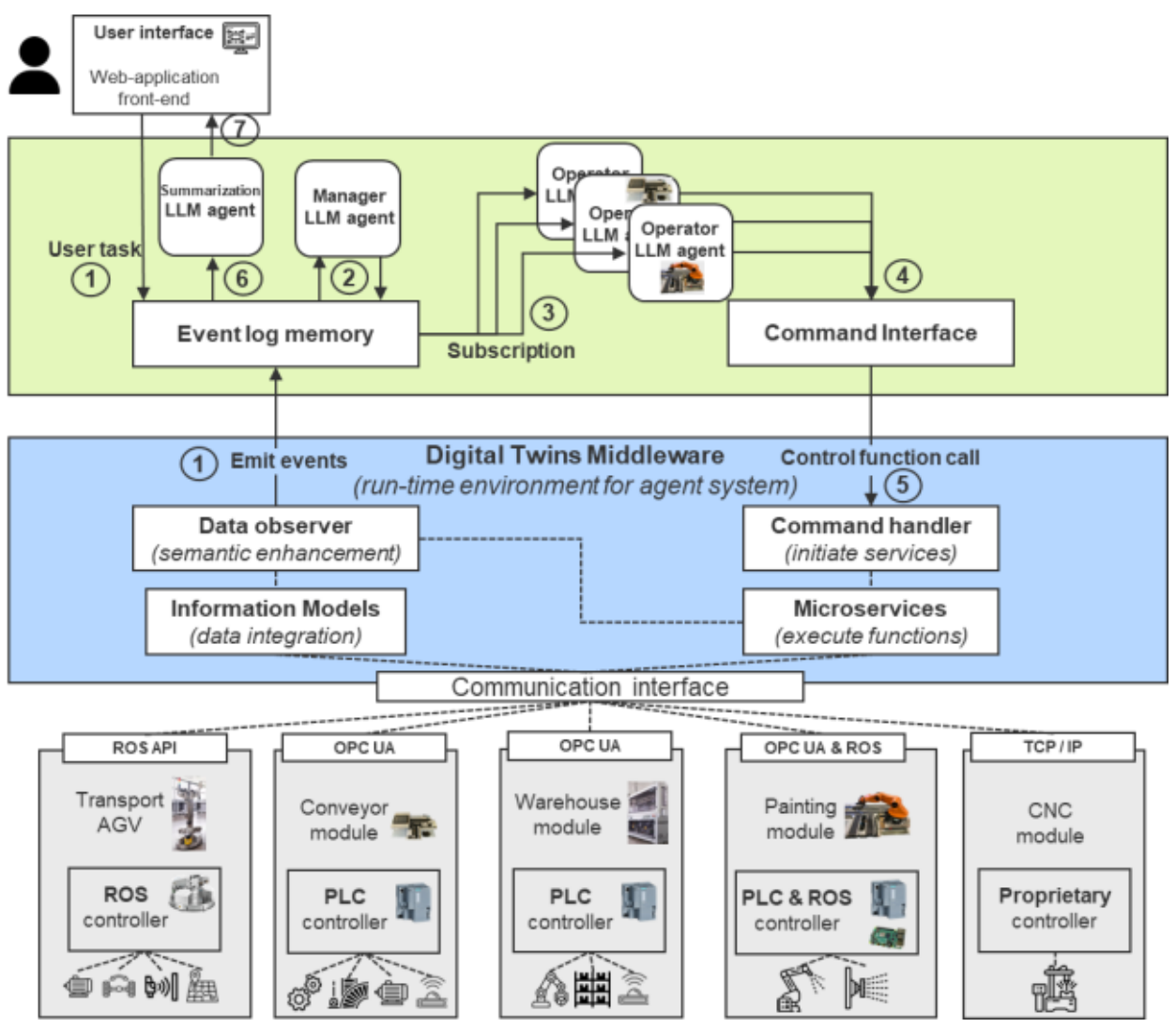

Figure 1 The overall system setup

## *B. Digital Twins and Semantics*

A digital twin is software that provides a synchronized digital representation of physical assets[5], [6]. It maintains information models that integrate field data and offers services to other software components. Overall, it serves as the foundation for a runtime environment supporting high-level applications, especially indispensable for LLM system.

Another critical aspect for integrating LLM in industrial applications is semantics. Systems operating at different levels often interpret the same data differently depending on the context[7]. For instance, "a bit flip from 0 to 1" in a PLC program indicates "motor on" at the field device level, "workpiece transport initiation" at a higher control level, and "logistics task starts" at the production planning level. These variations of semantics become more apparent when viewed through the lens of the automation pyramid, as illustrated in Figure *2*.

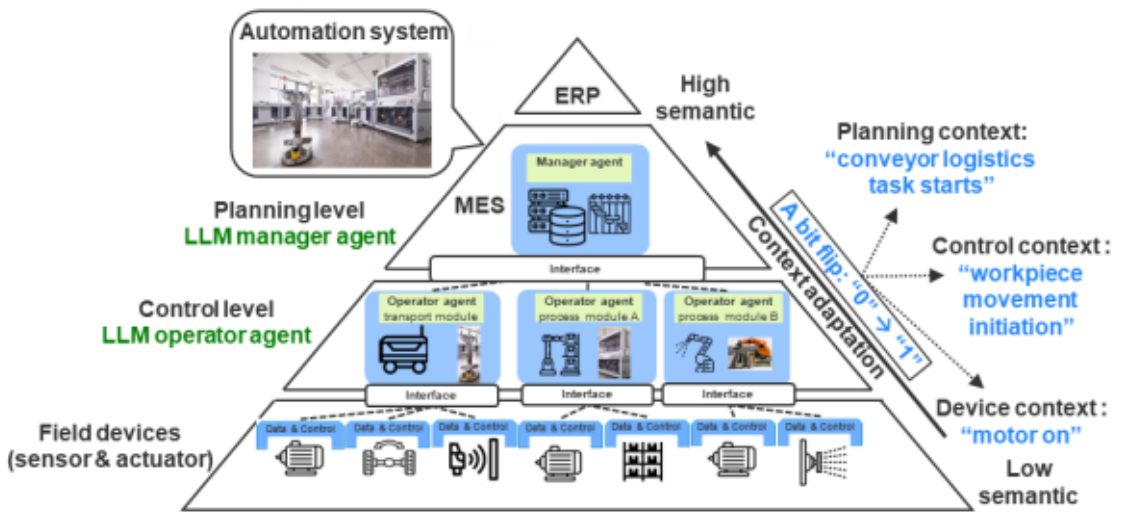

Figure 2 View of automation system modules and components from a hierarchy perspective: Automation Pyramid.

This necessitates a semantic enhancement process across various abstraction levels. We develop a data observer software component (refer to Figure *1*) for this purpose. The data observer monitors data in information models and converts them into textual expressions on different semantic abstraction levels. As different LLM-agents have distinct requirements for how data and changes should be interpreted in their specific task contexts, we pre-define the rules determining which events should be emitted on data changes and customize the text content for each agent, as shown in Figure *3*.

```
[Task Planner][Manager][12:04:23] task assigned: retrieve a 'white plastic cylinder' from the storage station.
[Storage Station][System][12:04:23] task received: retrieve a 'white plastic cylinder' from the storage station.
[Inspection Station][System][12:04:44] BG27 detects a workpiece at the outlet of conveyor C2.
[Storage Station][System][12:04:45] BG56 detects a carrier at the infeed of conveyor C1.
[Storage Station][Operator][12:04:45] Storage Station calls function: conveyor_1_run('forward', 13).
[Storage Station][Operator][12:04:45] Conveyor C1 starts running for 13 seconds.
[Storage Station][System][12:04:47] A carrier passes BG56.
[Storage Station][System][12:04:52] BG51 detects a carrier at the holder H2 on conveyor C1.
[Storage Station][Operator][12:04:53] Storage Station calls function: query_inventory_workpiece_position('white
 plastic cylinder').
[Storage Station][System][12:04:53] The 'white plastic cylinder' is located on shelf 'A_13'.
[Storage Station][Operator][12:04:54] Storage Station calls function: robot_arm_pick('A_13').
[Storage Station][System][12:04:54] Robot arm has started picking the workpiece from position A_13.
```

Figure 3 A snippet of the events subscribed by an LLM agent.

## *C. LLM Agent*

Recent studies on LLM applications have focused on the concept of LLM agents [8]. In the scope of this paper, an agent is defined as a software component that 1) is responsible for solving specific task requirements, 2) is associated with a physical asset and can be embodied in the form of an automation module.

Building on our previous work [9], [10], which introduced a hierarchical manager/operator LLM agent structure for automation system planning and control, this paper presents a refined and more scalable system design. A new fundamental component of this system design is an event log memory, combined with subscription and broadcasting mechanism that provides time-ordered information to agents, thereby keeping them informed about ongoing activities.

Based on the tasks, the system features three distinct agent roles:

- **Manager Agent:** This agent responds to user commands or events and generates an operational plan. It assigns subtasks to operator agents through the event log and monitors the plan execution. This agent type leverages the reasoning capability of LLMs in problem-solving and planning.
- **Operator Agent:** This agent executes tasks assigned by the manager agent or reacts to events by generating function call commands to control the ongoing production process. The operator agents are embodied with diverse automation modules to execute operations. This agent type leverages the reasoning and code understanding capabilities to control the operations within the automation system.
- **Summarization Agent:** This agent subscribes to the event log and provides a summary of system operations for the user. LLMs' long context understanding and alignment with human preferences are highly relevant for this task role. (cf. Figure *1*)

Each agent type has clearly defined responsibilities and interacts with the system according to its role using prompts.

## *D. Data-driven LLM Application and Testing*

To assess the performance of the agent system performing automation tasks, we utilize this setup to generate test datasets. These datasets comprise pairs of system-generated events and agent prompts alongside the expected outputs. This enables systematic testing of the LLMs' responses under various operational conditions.

Furthermore, the established framework enables the runtime environment for LLM agents through digital twins of the industrial automation system. If historical records of changes in the information model in the digital twin system are available, these can be utilized to develop datasets tailored for downstream automation and robotics tasks. This not only allows for testing of the LLMs but also facilitates the post-training of general LLMs to adapt them for the use case of controlling automation system. This agent system primarily processes textual data, where the effectiveness of the system hinges on the LLMs' ability to accurately interpret and reason with the digital twin system generated textual events.

In the following section, we further elaborate on the details of the proposed system for this application and provide a formal description of the framework's conceptual and structural composition.

## III. THE AGENT SYSTEM FRAMEWORK DESIGN

The organization of agent collaboration is crucial for effective application. For manufacturing process planning and control, we adopt a manager-operator model, applied across different abstraction levels following the automation pyramid (cf. Figure *2*). Additionally, we introduce a summarization agent to generate reports based on the event log for users, as shown in Figure *1*, though it is omitted in this section for brevity.

### A. Manager Agent and Operator Agent for Planning and Control

Our approach follows a manager-operator paradigm for task orchestration, as visualized in Figure 4, building on our previous works [9], [10]. The manager agent is a planning module responsible for processing user input tasks and decomposing them into sub-tasks to form a production plan. Operator agents are designed to control specific automation modules, receiving tasks from the manager and executing them accordingly.

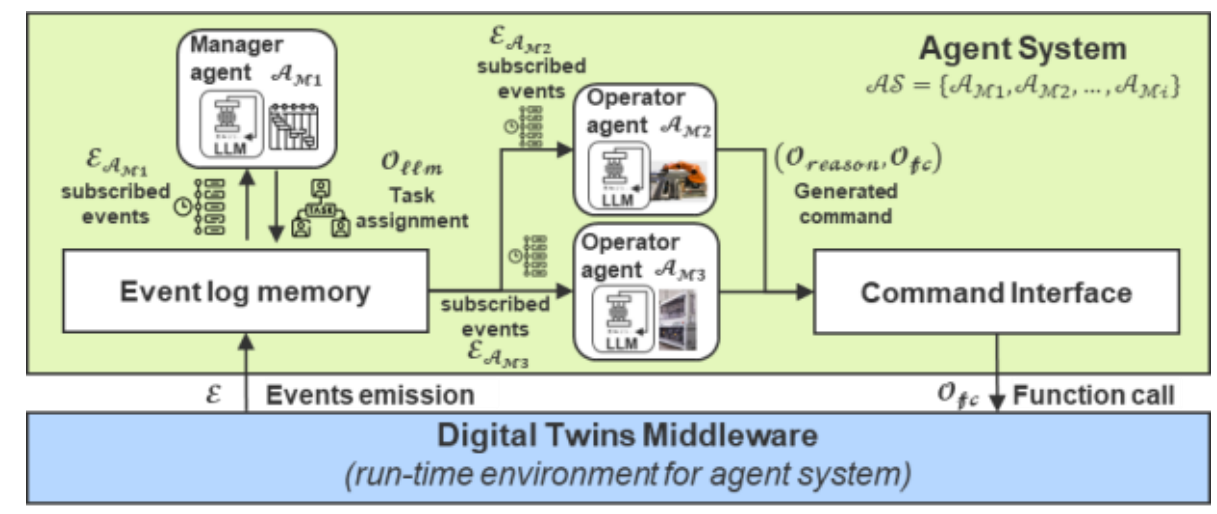

Figure 4 The agent system consisting of manager agent for planning and operator agents for controlling

To clearly define how our agent system operates, we describe its main concepts and their relationships. Our primary goal is to enable modular, flexible control over automation components by leveraging LLM-powered agents.

**The Agent System:** The overall system is controlled by several software agents, with each agent managing a dedicated automation module. This ensures responsibilities are cleanly separated and easy to scale.

**Automation Module:** Each **automation module** represents a specific piece of equipment or subsystem within the overall automation environment. An automation module is structured as follows:

- **Components**: The sensors, actuators, or other physical devices the agent can interact with.
- **Functions**: Operations that can be triggered on the module, such as starting a conveyor, picking a part, or processing a workpiece.

**Event Log and Subscriptions:** The entire system shares a global **event log**—a time-ordered list of all events emitted by the automation modules or agents. Every time the system perceives a change, or the system state updates, the corresponding event is added to this log with a timestamp.

**A subscription mechanism:** To remain focused and efficient, each agent only needs to be aware of the subset of events relevant to its role. This is achieved using a **subscription mechanism** and each agent subscribes to particular event types, filtering the global event log so it only observes changes within its specific area of responsibility.

**Prompt Construction and LLM Decision-making:** Whenever an agent needs to make a decision, it constructs a prompt that summarizes:

- Its current context (role and capabilities)
- The latest relevant events
- Applicable standard operating procedures (SOPs)
- A record of the current system state as perceived by that agent

This prompt is then submitted to an LLM, which processes the information and generates the output that contains:

- A **reasoning statement** explaining the rationale behind its intended action
- A **function call**—the concrete command that will be executed on the automation module

More details are described in TABLE I.

### B. Formal Description of the LLM Agent System

We provide a formal specification of the agent framework's conceptual and structural composition, as well as the relationships between the components, which underpin its software-technical implementation.

In the context of this paper, an LLM agent is a software component that processes textual data to control an automation module, with the LLM agent construct shown in Figure *5*.

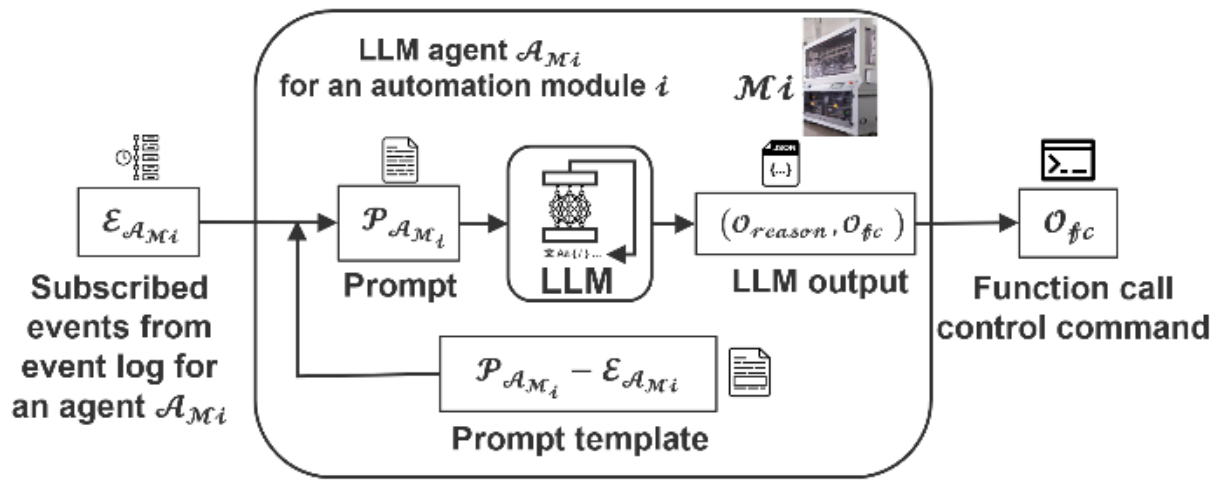

Figure 5 LLM agent processing cycle for command generation

*Agent System* $\boldsymbol{\mathcal{AS}}$:

$$\mathcal{AS} = \{\mathcal{A}_{\mathcal{M}1}, \mathcal{A}_{\mathcal{M}2}, \ldots, \mathcal{A}_{\mathcal{M}i}\}$$

An agent system $\mathcal{AS}$ consists of several agents, and each agent $\mathcal{A}_{\mathcal{M}i}$ (as illustrated in Figure *5*) is responsible for controlling a specific automation module $\mathcal{M}i$.

*Automation Module* $\boldsymbol{\mathcal{M}_i}$ :

$$\mathcal{M}_i = (\mathcal{C}_{\mathcal{M}_i}, \mathcal{F}_{\mathcal{M}_i}, \mathcal{E}_{\mathcal{M}_i})$$

Note that an overall automation system consists of multiple automation modules $\{\mathcal{M}_1, \mathcal{M}_2, \ldots, \mathcal{M}_i\}$.

Each individual automation module $\mathcal{M}_i$ comprises a tuple, where:

- $\mathcal{C}_{\mathcal{M}_i}$ is the set of components of the $\mathcal{M}_i$.
- $\mathcal{F}_{\mathcal{M}_i}$ is the set of functions exposed by $\mathcal{M}_i$.
- $\mathcal{E}_{\mathcal{M}_i}$ is the set of events that are in the scope of $\mathcal{M}_i$.

The individual events in set $\mathcal{E}_{\mathcal{M}_i}$ are pre-defined, automatically emitted upon state changes in the module. This events generation mechanism maps low-semantic data to high-semantic information for process planning and control.

*Event Log Memory $\boldsymbol{\mathcal{E}}$ that collects all the events:*

$$\mathcal{E} = \{(e_1, t_1), (e_2, t_2), \dots, (e_t, t_t) \mid t_1 < t_2 < \cdots < t_t\}$$

The event log memory records a sequence of **all** events $(e_t, t_t)$ ordered chronologically, representing the history of state changes in the automation system.

*Subscription Mechanism $\boldsymbol{\mathcal{S}}$:*

$$\mathcal{E}_{\mathcal{A}_{\mathcal{M}i}} = \mathcal{S}(\mathcal{A}_{\mathcal{M}i}) \subseteq \mathcal{E}$$

An agent $\mathcal{A}_{\mathcal{M}i}$ subscribes to the event log memory $\mathcal{E}$ to retrieve relevant events into its own event log memory $\mathcal{E}_{\mathcal{A}_{\mathcal{M}i}}$. $\mathcal{S}(\mathcal{A}_{\mathcal{M}i}) \subseteq \mathcal{E}$ denotes a selective function $\mathcal{S}$ that allows the agent $\mathcal{A}_{\mathcal{M}i}$ to subscribe to events from $\mathcal{E}$, thereby limiting the agent's scope of observation to relevant events.

*Agent Prompt $\boldsymbol{\mathcal{P}_{\mathcal{A}_{\mathcal{M}i}}}$:*

$$\mathcal{P}_{\mathcal{A}_{\mathcal{M}i}} = \text{Textual}\left(\mathcal{R}_{\mathcal{A}_{\mathcal{M}i}}, \mathcal{C}_{\mathcal{M}_i}, \mathcal{F}_{\mathcal{M}_i}, \mathcal{SOP}_{\mathcal{A}_{\mathcal{M}i}}, \mathcal{E}_{\mathcal{A}_{\mathcal{M}i}}\right)$$

A textual represented prompt $\mathcal{P}_{\mathcal{A}_{\mathcal{M}i}}$ for an agent $\mathcal{A}_{\mathcal{M}i}$ that integrates the elements listed in TABLE I.

*From LLM Generated Output $\boldsymbol{\mathcal{O}_{\ell\ell m}}$:*

$$\mathcal{O}_{\ell\ell m} = \left(\mathcal{O}_{reason}, \mathcal{O}_{fc}\right)$$

$\mathcal{O}_{\ell\ell m}$ consists of LLM's generated reasoning and a function call.

### C. The Prompting Method

The design of the prompt is pivotal in this framework. The prompt serves several purposes except as merely instruction for LLM agents, but also to 1) incorporate knowledge about the automation system, 2) serve as an integration point to connect the LLM with real-time events, and 3) later serves as prefix when training a LLM in a supervised fine-tuning manner, and, notably, tokens in the prompt are not included in the loss calculation. The construct of the prompt is described in TABLE I.

TABLE I. CONSTRUCT OF THE PROMPT

| **Prompt Section** | ***Definition*** | ***Typical Examples*** [a] |
|---|---|---|
| **R**ole Definition $\mathcal{R}_{\mathcal{A}_{\mathcal{M}i}}$ | Describe the role and the task responsibility of the agent. | You are the operator of an automation module called "Storage Station", responsible for handling of workpieces and directing material transport on conveyors. |
| **C**omponent Description $\mathcal{C}_{\mathcal{M}_i}$ | Entries of component description about the sensors and actuators. | - BG56 is a proximity sensor located at one end of the Conveyor C1… <br>- H1 is a holder located in the middle of the Conveyor C2 at the export verification point, Holder H1 can hold the workpiece in position. <br>- TF81 is an RFID sensor located in the middle of the Conveyor C1 and at Holder H2 of the pick and place point; it can read the workpiece information. |
| Callable **F**unction $\mathcal{F}_{\mathcal{M}_i}$ | The parametrized functions provided by digital twins middleware that can be invoked by the agent. | - conveyor_1_run(direction, time) <br>- This function is used to start Conveyor C1 and run it in a specified direction for a specified duration. |
| **S**tandard **O**peration **P**rocedure $\mathcal{SOP}_{\mathcal{A}_{\mathcal{M}i}}$ | Specify the behavior of the agent under normal operation. | - After detecting a carrier at the entrance, transport the carrier to the pick and place point. <br>- When the carrier arrives at the pick and place point, query the workpiece position in the storage. |
| Auxiliary Instruction | Other instructions for guiding LLM to generate output with desiered format. | - A series of events will provide you the information about the current state of the system. <br>- You should follow the following input and output pattern to generate your response in JSON format. First provide a short reason and then generate a function call. |
| **E**vent Log Input $\mathcal{E}_{\mathcal{A}_{\mathcal{M}i}}$ | The dynamic information that require an agent to react. | [Manager][12:04:23] task assigned: retrieve a 'white plastic cylinder' from the storage station. <br>[System][12:04:23] task received: retrieve a 'white plastic cylinder' from the storage station. <br>[System][12:04:23] BG56 detects a carrier at the infeed of conveyor C1. |
| **O**utput (to be generated) $\mathcal{O}_{\ell\ell m}$ | The generated command by the agent to control the system. | { "reason": "Carrier detected at entrance, initiate transport to pick and place point", "command": "conveyor_1_run('forward', 13)"} |

[a] *a*. more comprehensive examples can be accessed from ***GitHub***

Next, we provide a detailed explanation of several special components within this prompt design.

#### *1) Event log*

This design of the event log is driven by the rationale that **time** and **information** are indispensable in control and planning tasks. The event log provides the LLM with dynamic information about production operations in a time-sequential order, organized as **events**. Based on our consideration and evaluation, this is a succinct way to represent the required information in text form. The first label indicates the scope of the message subscription, the second label indicates the source of the message, and the third label is a timestamp, accompanied by a textual description of the event occurring at that moment, as illustrated in Figure *6*.

```
[Storage Station][System][12:05:00] BG26 detects a workpiece at the infeed of the conveyor C2.
[Storage Station][Operator][12:05:01] Storage Station calls function: conveyor_2_run('forward', 13).
[Storage Station][Operator][12:05:01] Conveyor C2 starts running for 13 seconds.
[Storage Station][System][12:05:03] A workpiece passes BG26.
[Storage Station][System][12:05:07] BG21 detects a workpiece at the Holder H1 on conveyor C2.
[Storage Station][Operator][12:05:07] Storage Station calls function: export_verify().
[Storage Station][System][12:05:07] This workpiece is verified to export from the storage station.
[Storage Station][Operator][12:05:08] Storage Station calls function: conveyor_2_run('forward', 8).
[Storage Station][Operator][12:05:08] Conveyor C2 starts running for 8 seconds.
[Storage Station][Operator][12:05:09] Storage Station calls function: H1_release().
[Storage Station][System][12:05:09] Holder H1 is released.
[Storage Station][System][12:05:10] A workpiece passes BG21.
```

Figure 6 A snippet of the subscribed events from the digital twin middleware for "Storage Station".

#### *2) Event emission from digital twin middleware*

An event emission is automatically triggered upon state changes in the information models, messages from other agents, or the initiation or completion of an operation. These event descriptions are pre-defined in the data observer and use natural language to describe high-level semantic information about system activities. The emitted events are recorded in the event log and are then consumed by the LLM agents according to their subscription scopes, allowing the agents to generate appropriate commands in response.

#### *3) Standard Operation Procedure (SOP)*

In manufacturing, Standard Operating Procedure (SOP) provides guidelines for consistent and safe operation. In our approach, we repurpose this concept of "SOP" to refer to instructions for the LLM on how to respond to specific conditions during machine operation.

Similar to how traditional SOPs guide human operators, the SOP in the prompt informs the LLM about standard protocols for operating machines in specified situations.

In contrast to "n-shot" prompting methods , this SOP-based approach enables the configuration of LLMs at a higher level of abstraction and allows for integration of behavioral knowledge. From the perspective of the user who needs to modify system behaviors, this approach facilitates programming the system using natural language by specifying SOP for LLM agents.

For situations that are not considered in SOP, the LLM agents usually, according to our experiment observation, generate commands based on common sense, enabling them to handle scenarios flexibly based on the information given in the prompt, though the responses are not always accurate.

#### *4) Reason in LLM output*

Before generating a command, an LLM is instructed to first generate a reason. This serves three main purposes: First, this design mirrors the Reflection-Act (ReAct) process [11] in a minimalist manner. Second, it enhances transparency and explainability, allowing for deeper evaluation by comparing the generated reason with the reference reason. Third, reference reasons in the dataset can provide additional knowledge during fine-tuning and enable calculating loss and weight update on more data, helping the LLM to learn how and why specific planning and control decisions are made.

## IV. DATASET CREATION

In this section, we introduce how we create and organize the dataset for testing and training the LLM for this downstream use case, as illustrated in Figure *7*.

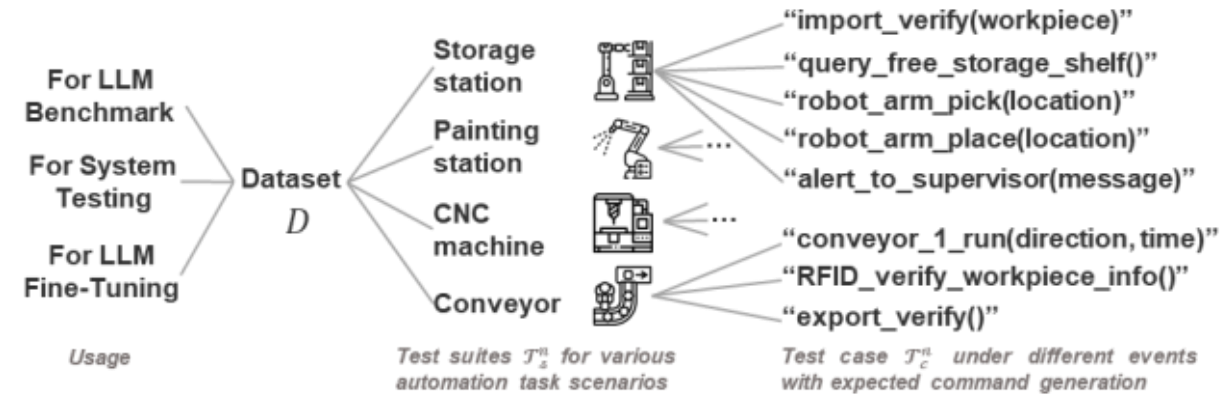

Figure 7 The construct of the dataset for testing and training

### *A. Dataset Creation Based on the Agents System and Prompting Method*

The LLM agents generate function calls based on the information provided in the prompt. During operation, the underlying digital twin system automatically emits new events in the event log. The agent is continuously updated with real-time textual information to generate a response, which includes a function call to control the equipment and a brief explanation for the reason.

### *B. The Dataset*

To systematically evaluate and further improve our LLM-based automation system, we construct a structured dataset that supports both **testing** and **fine-tuning**. The hierarchy and components of this dataset are as follows:

- **Test Point**: A *test point* represents a single step in agent interaction. It consists of the agent's prompt (which includes its current role, available components, callable functions, standard operating procedures, and event history) together with a set of newly observed events. Each test point essentially captures the context in which the agent must make a decision.
- **Test Case**: A *test case* pairs a test point with the corresponding reference output from a human or documented ground truth. This output includes: (1) the function call that should be executed in response, and (2) a brief reasoning explaining the choice. In other words, each test case presents the LLM with a situation and the correct response.
- **Test Suite**: Several test cases that focus on a particular operational scenario are grouped into a *test suite*. Each test suite covers all the typical and unexpected events relevant to a specific workflow, such as inventory management or anomaly handling.
- **Dataset**: The *complete dataset* is made up of multiple test suites, thus covering a wide range of task scenarios and variations. This dataset is used not only to quantify the system's accuracy, but also to serve as a resource for further LLM training.

Importantly, this dataset has dual purposes. First, it enables systematic testing—allowing us to benchmark model performance across a variety of operational situations. Second, as highlighted in Figure *8*, it can be directly used as training data for supervised fine-tuning of the LLM. By learning from these labeled cases, the LLM can better adapt to the specific operation sequence of the automated process.

### *C. Formal Description of the Dataset*

A formal specification of the conceptual and structural composition that underpins the software technical implementation is provided as follows:

*Test Point* $\boldsymbol{\mathcal{T}_p}$:

$$\mathcal{T}_p = \left(\mathcal{P}_{\mathcal{A}_{\mathcal{M}i}} \cup \{\mathcal{E}_{\mathcal{A}_{\mathcal{M}i},[t+1,t+k]}\}\right)$$

A test point $\mathcal{T}_p$ is defined as an agent prompt $\mathcal{P}_{\mathcal{A}_{\mathcal{M}i}}$ containing incremental $k$ new events $\mathcal{E}_{\mathcal{A}_{\mathcal{M}i},[t+1,t+k]}$ for testing the output of an agent $\mathcal{A}_{\mathcal{M}i}$. When expanded:

$$\mathcal{T}_p = \text{Textual}\left(\mathcal{R}_{\mathcal{A}_{\mathcal{M}i}}, \mathcal{C}_{\mathcal{M}_i}, \mathcal{F}_{\mathcal{M}_i}, \mathcal{SOP}_{\mathcal{A}_{\mathcal{M}i}}, \mathcal{E}_{\mathcal{A}_{\mathcal{M}i},[0,t]} \cup \mathcal{E}_{\mathcal{A}_{\mathcal{M}i},[t+1,t+k]}\right)$$

*Test Case* $\boldsymbol{\mathcal{T}_c}$:

$$\mathcal{T}_c = \left(\mathcal{T}_p, \mathcal{O}_{\ell\ell m}^*\right)$$

A test case is the combination of a test point $\mathcal{T}_p$ and the expected reference output $\mathcal{O}_{\ell\ell m}^*$ from the LLM, where $\mathcal{O}_{\ell\ell m}^*$ consists of the expected function call to be generated and a reference reason:

$$\mathcal{O}_{\ell\ell m}^* = \left(\mathcal{O}_{reason}^*, \mathcal{O}_{fc}^*\right)$$

*Test Suite* $\boldsymbol{\mathcal{T}_s}$:

$$\mathcal{T}_s = \{\mathcal{T}_c^1, \mathcal{T}_c^2, \ldots, \mathcal{T}_c^n\}$$

Test cases $\mathcal{T}_c^1$ to $\mathcal{T}_c^n$ are organized into test suite $\mathcal{T}_s$, each corresponding to a specific operational task scenario.

*Dataset* $\boldsymbol{\mathcal{D}}$:

$$\mathcal{D} = \{\mathcal{T}_s^1, \mathcal{T}_s^2, \ldots, \mathcal{T}_s^n\}$$

The complete dataset $\mathcal{D}$ consists of a collection of test suites from $\mathcal{T}_s^1$ to $\mathcal{T}_s^n$, representing various task scenarios. The dataset is used for both testing and fine-tuning the LLM for the application.

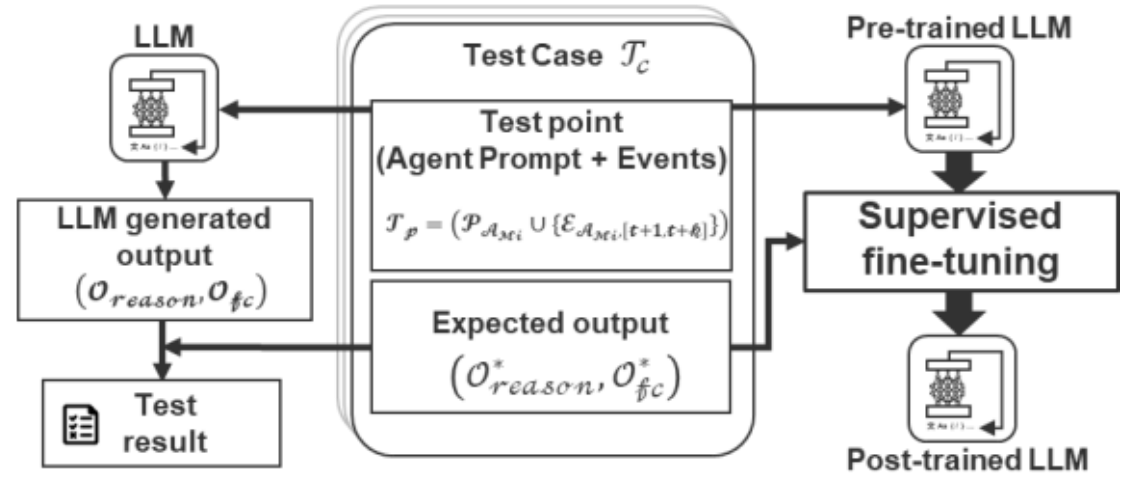

Figure 8 Two uses of the created dataset: Testing and Post-training

## V. EXPERIMENTS

Using the created dataset, we first evaluate several open-source pre-trained and a proprietary model GPT-4o. Then, we applied supervised fine-tuning to further train these models on the created dataset.

The objectives are twofold: firstly, to evaluate the fine-tuning enhancement of the LLM agent's performance on this specific task; and secondly, to gain insight into the cost-performance trade-offs involved.

### A. *Test case creation*

In contrast to normal operation where LLM agents observe events before deciding on machine operation, dataset creation involves a reverse process, as depicted in Figure *9*.

In this case, the dataset is created without LLM agents, but with direct user input. With a specific task in mind, the user manually operates the command interface to interact with the automation system. As this process unfolds, the digital twin system automatically generates and records relevant events. The user finally provides a description of the intended task process. This approach captures the three essential elements for dataset: the events, the command calls, and the initial user task request. The dataset contains the knowledge necessary for successful execution of intended tasks.

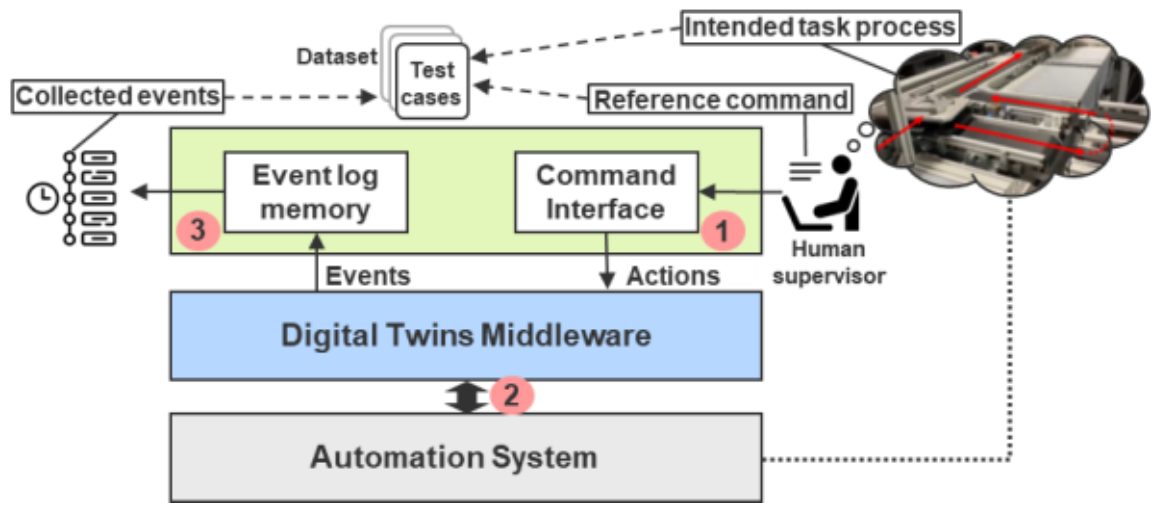

Figure 9 Dataset creation process

Using this approach, we create various task scenarios for handling typical situations in factory operations, such as processing user orders, responding to spontaneous events, or handling abnormalities. Our initial collected dataset contains 100 test cases $\mathcal{T}_c = (\mathcal{T}_p, O^*_{\ell\ell m})$, each consisting of a complete prompt and the expected LLM output. This dataset can be used to assess the performance of pre-trained LLMs in controlling industrial automation systems.

In our initial dataset, 68% of the control command generation involves repetitive tasks in our initially prepared dataset, where the LLM agents should follow the SOP routines. The remaining 32% consist of non-routine tasks, requiring LLM agents to respond to unprecedented events through autonomous decision-making.

### B. *Metrics for Evaluation*

We apply two metrics to evaluate the system's performance.

- **Correctness Rate:** Measures whether the generated command matches the reference command in the dataset.

- **Reason Plausibility:** Assessed through human evaluation, using a Likert scale from 1 to 5 to rate the plausibility of the generated reason.

These two metrics together provide a more granular metric to identify loss. In some cases, the command is incorrect while the reason may be plausible.

### C. *Special requirements of automation tasks*

Given the nature of industrial automation, there are two main aspects of the requirements:

- **Accuracy in Repetitive Operations:** Industrial automation tasks often require 100% accuracy in repetitive operations. This repeatability implies that some routine tasks can be anticipated. This requirement can be evaluated by assessing whether the LLM agents follow the SOP to successfully complete the tests, or whether the model improves by learning from the dataset to perform in-sample tasks.

- **Handling Unexpected Events:** The system shall be capable of responding to unexpected events not predefined in the SOP or present in the training dataset. It should demonstrate generalization by flexibly handling unforeseen situations that were not anticipated during development. The language model should use its learned knowledge to generate appropriate responses to spontaneous events—an ability typically lacking in traditional automation systems.

Based on these considerations, we perform a comprehensive evaluation and model fine-tuning.

### D. *The evaluation*

We use different LLMs as inference engines to power the agent systems. GPT-4o is selected to represent the state-of-the-art performance achievable by LLMs, while other open-source models are chosen to represent those that can be practically deployed in on-premises industrial settings. To account for the trade-off between model complexity and performance, we compare larger models (70B range) with smaller ones (7B range). We chose GPT-4o[2], Llama3 models[12], Qwen2 models[13] and Mistral models[14] for our experiments.

TABLE II. EVALUATION OF PRE-TRAINED AND FINE-TUNED LLMS

| | **Evaluation based on 100 test points*** ** | | | | | | |
|---|---|---|---|---|---|---|---|
| | ***GPT-4o*** | **Llama-3-70B-Instruct** | **Llama-3-8B-Instruct** | **Qwen2-72B-Instruct** | **Qwen2-7B-Instruct** | ***Mistral-7Bx8-Instruct-v0.2*** | ***Mistral-7B-Instruct-v0.2*** |
| Pre-trained (all) | **81% \| 4.7** | **75% \| 4.3** | **37% \| 2.8** | **70% \| 4.0** | **65% \| 3.7** | **29% \| 2.4** | **45% \| 2.9** |
| Pre-trained (SOP) | 100% \| 5.0 | 87% \| 4.5 | 53% \| 3.1 | 85% \| 4.5 | 63% \| 3.6 | 34% \| 2.4 | 37% \| 2.5 |
| Pre-trained (Unexpected) | 41% \| 4.0 | 50% \| 3.8 | 3% \| 2.2 | 38% \| 3.0 | 69% \| 4.0 | 19% \| 2.3 | 63% \| 3.7 |
| SFT (all) | **100% \| 5.0** | **95% \| 4.8** | **96% \| 4.9** | *** 66% \| 3.9** | **97% \| 4.9** | **45% \| 3.1** | **** N.A.** |
| SFT (SOP) | 100% \| 5.0 | 94% \| 4.8 | 99% \| 4.9 | * 82% \| 4.4 | 97% \| 4.9 | 61% \| 3.6 | ** N.A. |
| SFT (Unexpected) | 100% \| 5.0 | 97% \| 4.9 | 91% \| 4.7 | * 31% \| 2.8 | 97% \| 5.0 | 9% \| 2.3 | ** N.A. |

Value: (accuracy of their generated commands)% | (averaged reason plausibility 1-5)

* : we ran out of GPU capacity and used LoRA instead of full fine-tuning for the Qwen2-72B-model.

** N.A.: Our full-fine-tuning made Mistral-7B model unstable, and it generated unusable echo texts.

***: Details about the SFT, dataset and evaluation sheets on GitHub: YuchenXia/LLM4IAS

#### 1) *Evaluation of pre-trained LLM*

We begin by evaluating the original pre-trained models. In automation tasks, 1) some are routine processes where the LLM agent can follow **SOP** guidelines in agent prompts to operate the automation system, while 2) others require the agent to autonomously respond to **unexpected** events, for which reactions have not been instructed in agent prompts. We distinguish between these two types of tasks in our evaluation.

Based on the evaluation results, GPT-4 generally outperforms other open-source models in interpreting agent prompts and events to generate control commands, though their performance varies significantly. Each model also exhibits distinct "personalities" in this use case.

*2) Evaluation of post-trained LLM based on created dataset*

Using the collected dataset, we apply supervised fine-tuning (**SFT**) to assess how training open-source models can improve the LLM's performance for this specific downstream task. This training has the potential to enable the customization of a general LLM for intelligent control of specialized automation equipment. For GPT-4o, we used OpenAI's proprietary fine-tuning API to explore the capabilities that LLMs can achieve, even though the training methods may vary.

For two main considerations, we train the models on the fixed dataset created during this research project: 1) the machine is designed to execute specialized tasks most of time repetitively, and 2) the goal of fine-tuning is to evaluate whether the model can effectively learn the machine's operational knowledge. Additionally, in one of our concurrent research experiments, we observed that **larger LLMs** generally exhibit **less significant** overfitting or catastrophic forgetting than **smaller LLM** during a k-fold cross-validation fine-tuning based on limited results for this use case. However, we will address this interesting hypothetical finding in the future as we work to overcome cost constraints and dataset scarcity issue.

OpenAI's model and fine-tuning services outperform other models, and the GPT-4o model quickly learns from the samples how to control the automation systems. Other models also demonstrated reasonably good performance. Interestingly, fine-tuned smaller LLMs did not necessarily underperform in this particular use case. However, our contingency LoRA[15] fine-tuning yielded poor results in our experiments and led to a decrease in model performance.

An accuracy of less than 100% means that the results are not fully reliable for directly controlling an automation system, as it could result in stops and errors during operation. However, they can be developed as an assistant system, proposing next actions for human supervisors to approve in a human-machine collaboration use case scenarios.

## VI. Conclusion

In this paper, we introduce a novel application-oriented framework that enables the use of LLMs to control industrial automation systems. For system developers, the development can be divided into two phases: 1) modularizing the system and creating interoperable interfaces to establish the physical and digital foundation for agents, and 2) creating datasets and applying LLM-specific prompting and training methods. The result is an end-to-end solution that allows an LLM to control automation systems, with the reconfiguration process and human machine interactions made more intuitive through natural language. We are continuing to refine our design and implementation to further increase the technology readiness level and will post new results from this ongoing research.

## Acknowledgment

This work was supported by Stiftung der Deutschen Wirtschaft (SDW) and the Ministry of Science, Research and the Arts of the State of Baden-Wuerttemberg within the support of the projects of the Exzellenzinitiative II.